# AN INTERACTIVE DASHBOARD FOR REAL-TIME ANALYTICS AND MONITORING OF COVID-19 OUTBREAK IN INDIA: A PROOF OF CONCEPT

Arun Mitra, Achutha Menon Centre for Health Science Studies, SCTIMST, India , dr.arunmitra @gmail.com

Biju Soman, Achutha Menon Centre for Health Science Studies, SCTIMST, India , bijusoman@sctimst.ac.in

Gurpreet Singh, Achutha Menon Centre for Health Science Studies, SCTIMST, India, drgurpreet_amc@sctimst.ac.in

**Abstract:** Data analysis and visualization are essential for exploring and communicating findings in medical research, especially in epidemiological surveillance. Data on COVID-19 diagnosed cases and mortality, from crowdsourced website COVID-19 India Tracker, Census 2011, and Google Mobility reports have been used to develop a real-time analytics and monitoring system for the COVID-19 outbreak in India. We have developed a dashboard application for data visualization and analysis of several indicators to follow the SARS-CoV-2 epidemic using data science techniques. A district-level tool for basic epidemiological surveillance, in an interactive and user-friendly manner which includes time trends, epidemic curves, key epidemiological parameters such as growth rate, doubling time, and effective reproduction number have been estimated. This demonstrates the application of data science methods and epidemiological techniques in public health decision-making while addressing the gap of timely and reliable decision aiding tools.

## INTRODUCTION

As of 15th May 2021, over 160 million confirmed coronavirus infections with over three million deaths across 192 countries were reported worldwide (Dong et al., 2020; *World Health Organization Coronavirus (COVID-19) Dashboard*, 2021). According to the WHO COVID-19 situation update report on 11[th] May 2021, India had over 22 million confirmed COVID-19 cases (3 million active cases) with close to 250,000 deaths (Ministry of Health and Family Welfare (MoHFW), India, 2021; World Health Organization & others, 2021). The government's decision of a nationwide lockdown during the initial phase of the epidemic and its sequential extension resulted in a significant reduction in the total number of COVID-19 cases in India. The effect of lockdown is also reflected by the decrease in the key epidemiological parameters of the COVID-19 outbreak in India(Chatterjee et al., 2020; Das, 2020; Mitra et al., 2020). However, this decrease was variable across different states and districts, which might call for a more cautious approach in the post-lockdown era (Basu et al., 2020). The variation across different geographical and administrative regions also highlights the need for a more tailored approach based on individual clusters and the epidemic phase rather than taking a blanket approach.

The rapid spread of the COVID-19 pandemic has demanded collective action against coronavirus disease on an unprecedented scale (Holmes et al., 2020; Lin et al., 2020; Ohannessian et al., 2020; Wong et al., 2020; World Health Organization & others, 2020b; Zhang et al., 2020). Governments, organizations, institutions, and individuals came up with a vast array of innovations and technological advances in the field of infectious disease epidemiology and vaccine research (Chesbrough, 2020; Chick et al., 2020; Guest et al., 2020; Le et al., 2020; World Health Organization & others, 2020a; Yamey et al., 2020). Some of these innovations include innovative tools such as crowdsourced data generation, apps, dashboards, and tracker websites to monitor and track the COVID-19 outbreaks (Boulos & Geraghty, 2020; Calvo et al., 2020; Carvalho et al., 2020; Chen et al., 2020; Dong et al., 2020). Dashboards quickly became a popular tool to visualize and monitor the coronavirus disease spread and aid in decision-making and implementation of containment measures by tracking the critical epidemiological parameters and studying the transmission dynamics in real-time. A Google trend analysis shows that the keyword "COVID dashboard" quickly rose to peak popularity by the end of March 2020, with many such dashboards with varying degrees of granularity and interactivity developed for various purposes being available online.

## 1.1. Dashboards for public health decision making

From the general public to policymakers, government agencies, and even corporate to research firms, many actively seek authentic information on COVID-19's impact on human lives — to stay updated and avoid misinformation-induced panic. Dashboards are tools that help in aggregating data from different sources in real-time and helps in decision making. They are tools to induce transparency in decision-making and explain the policy decision (Head, 2020). Another purpose of dashboards in COVID-19 has been to generate awareness about the case trends, tracking and monitoring the epidemics while guiding in corrective action in a timely fashion.

The applications and use of dashboards in the COVID-19 have been very diverse. These range from monitoring and mitigating preventive measures like social distancing and wearing masks to monitoring clinical trials. In the USA, dashboards were used to offer interactive visualization about the infection rates across different geospatial regions and the use of morgues in these areas. The authors used this tool for both aiding and increasing situational awareness as well as for capacity-building. It also helped the policy decision-makers and program managers to be better prepared for the epidemic and better utilize the available resources to the maximum benefit (Kaul et al., 2020). Dashboards also have been used in monitoring hospital infections and bed occupancy rates in the COVID-19 context as well as aiding in the monitoring of multiple drug and intervention trials (Thorlund et al., 2020). In the Democratic Republic of Congo, a dashboard was created based on the variables collected from street passers about the perceived utility of preventive measures, actual practices, and adherence to mask. This was found to be an effective visual stimulation for daily monitoring (Wimba et al., 2020). The team behind a US dashboard (WATCHERS) argues the need to report trends at the municipality level in order to have a better understanding of the ground reality, which was lacking in the Centre for Disease Control and Prevention reports (Wissel et al., 2020).

## 1.2. Indian context

There have been several dashboards developed by the Indian government and various state governments (*MyGov*, 2021; *World Health Organization Coronavirus (COVID-19) Dashboard*, 2021). A few dashboards for India are developed by the academic institutes (Harvard T.H. Chan School of Public Health, 2020; IIT Delhi, 2021) and by and volunteers (*Coronavirus in India*, 2021; *Covidtracking.In*, 2021). Table 1 shows a summary of some of the dashboards relating to COVID-19 in India. Apart from these, there are many other *Github* repositories, Tableau visualizations, and ArcGis pages that provide many insights on the transmission and impact of the COVID-19 epidemic. Crowdsourced initiatives like WhatsApp working groups and Telegram Folding groups that volunteer to maintain the tracking websites also play a crucial role. One such initiative is a tracking website (covid19india.org) curates count data published in state bulletins and official handles and provides an API for public use of the data. Many research papers, blog posts, visualizations and dashboards use these APIs for fetching the data. This is an excellent example of the benefits of information and communication technology for public benefit. Also, there have been instances where state governments realized the importance of dashboards for monitoring and tracking the COVID-19 epidemic is an illustration for the use of information technology in health governance (ETGovernment, 2020).

Though many dashboards have been developed with varying levels of granularity for COVID-19, most of the dashboards fail to provide the crucial insights necessary to understand the transmission phenomenon. Barring a few, the majority of the dashboards are descriptive and do not provide additional information beyond case counts and deaths. Some dashboards which provided projections were far from the observed cases. One of the significant limitations of many such dashboards was the lack of domain knowledge. Thus, none of these dashboards have integrated the epidemiological aspects with social-geographical and macro-environmental aspects and have not seen the phenomenon with background susceptibility. Decision-making in public health could be viewed as a complex network of actors, agencies, and technologies with dynamic relationships. Lots of factors influence the decision-making in public health and these can be viewed at different levels (individual level i.e., patient, caregiver, clinician, program manager, decision-maker, policymaker, politician etc.; population-level i.e., sociodemographic variations, geospatial differences, etc.; infrastructure level i.e., health system capacity, health information infrastructure, digitalization of healthcare etc.; policy level i.e., health information use policy, public trust, vision, implementation etc.). This process is very complex and based on many interactions and relationships. Having good quality health information and robust health information systems could enhance the decision-making process. The current COVID-19 pandemic makes the case stronger for the urgent need for a more efficient epidemic tracking mechanism that can provide insights at the district level while also being scalable and reproducible.

# OBJECTIVE

In this manuscript, we attempt to demonstrate the application of data analytics, data visualization, and epidemiology to develop an intuitive and interactive dashboard for real-time analytics and monitoring the COVID-19 outbreak in India based on crowdsourced data.

# MATERIALS AND METHODS

### 3.1. Data Source

In India, confirmed cases of COVID-19 and deaths are reported to the Ministry of Health and Family Welfare (MoHFW), Government of India, through the national reporting network. The health authorities of the state simultaneously publish daily bulletins of the same. The crowdsourced database and website maintained at https://covid19india.org curates these bulletins and publishes this data into an application programming interface (API) available for public use. The API includes district-level, state-level, and national-level datasets. The downloadable dataset is updated regularly and contains the latest available public data on COVID-19 in India. The source of the data relating to district level population density was Census 2011, conducted by the Government of India. A compilation of these figures is available from the Wikipedia page, which was scraped using *rvest* package for scraping hypertext markup language (HTML) and eXtensible markup language (XML) tables. Additional data sources include the Google Mobility Trends data (Google, 2021).

### 3.2. Variables used in the study

The variables used in the study are the daily caseload, including the confirmed cases, recovered, and deceased. These variables are provided by the API at https://covid19india.org. The population projections for 2020 for each state which were required to estimate indicators such as cases per million, deaths per million, and other relevant indicators such as for testing and vaccinations, were taken from the website of the Unique Identification Authority of India (UIDAI), which is the statutory body established under the Aadhaar Act, 2016 by the Govt of India (Government of India, 2021). Variables on COVID-19 vaccinations are sourced from the Open API provided by the COWIN portal, which is the official centralized portal for COVID-19 vaccinations in India. Additional variables reflecting the population dynamics, such as the mobility trends at the district level, were taken from the Google COVID-19 Community Mobility Report (Aktay et al., 2020; Google, 2021), while variables providing details on contact tracing, testing, and hospitalizations for the state of Kerala were obtained from the Department of Health Services, Kerala website. These data are provided as part of the daily situation report in the form of a PDF (portable document format) which is then scraped for data using packages such as *rvest* (Wickham & Wickham, 2016).

### 3.3. Comparison of Epidemiological Parameters of the First and Second Wave

The epidemiological parameters that were estimated were growth rate, doubling time, and effective reproduction number. A detailed description of the methods on the estimation is provided elsewhere (Mitra et al., 2020). The end of the first wave and the start of the second wave has been defined as the day with the least number of cases after the peak of the first wave. The peak of the first wave was estimated to be $19^{th}$ September 2020 (95% CI $18^{th}$ September – $21^{st}$ September). The valley was estimated to be $13^{th}$ February 2021. The growth rate and doubling time to the peak of the first wave were compared with the growth rate and doubling time to the peak of the second wave.

### 3.4. Projections and Validation

For the purpose of the projections of the daily incident cases, we have made some empirical decisions based on some considerations. Taking into consideration the second wave of COVID-19 in India, we thought the best time to freeze the dataset would be $1^{st}$ May 2021. This was done to include the most recent data but also to avoid any technical or erroneous entries in the recent past, which are usually rectified in a few days. We also decided the number of days for the projection of future incidence to be 15 as any projection beyond two weeks would mean a wide confidence interval and high error rate. Also, as the model is based on the regression of log incidence over time, it would be subject to overestimation as time progresses. We also decided on a 14-day moving average of the daily incident cases for modeling the future projections. For the epidemic simulation, we assumed that the branching process of the probability mass function follows a Poisson distribution.

Based on the considerations provided above, we then divided the dataset into two based on the date, $15^{th}$ April 2021. The cases before $15^{th}$ April 2021 were used to estimate the effective reproduction number by the time-dependent method developed by Wallinga and Teunis and improved by Cauchemez and colleagues (Cauchemez et al., 2006; Wallinga & Teunis, 2004). The effective reproduction number of $15^{th}$ April 2021 was then extracted and provided as a hyperparameter for the epidemic simulation along with the distribution of the serial interval. As already described, we projected for each state the daily incident cases for the next 15 days and plotted against the observed 14-day moving average to check for the robustness of the effective reproduction number and the validity of the projections.

### 3.5. Implementation

The dashboard was developed using R, version 4.0.4, a free and open-source statistical software, and the RStudio, version 1.4.1106, an IDE interface. Key packages used for the dashboard development were the *tidyverse* packages, *flexdashboard*, and *shiny*. Graphing libraries like *ggplot2* and *plotly* were used to create interactive and 3D data visualizations. Epidemiological analysis was done using the packages, *incidence*, *projections*, *R0*, and *EpiEstim*. Packages *rvest* and *fuzzyjoin* were used to scrape sociodemographic information from the web and enable data linkage. The packages *furr* and *future* were used to optimize the code execution.

We used the package *incidence* to model the incidence and estimate growth rate and doubling time and package *R0* to estimate the time-varying reproduction number (Rt) for different districts. The growth rate and doubling time were estimated by fitting an exponential model to the incidence data in the form of *log(y) = r \*t + b*; where *y* is the incidence, *t* is the time (in days), and r is the growth rate while b is the intercept or origin. The doubling time is then estimated by dividing the natural logarithm of two with the growth rate of the epidemic i.e., doubling time *(d) = log (2) / r*. The package *projections* was used to simulate the epidemic outbreaks and project their respective trajectories based on the state-specific transmission parameters. Regression of log-incidence overtime was used to model the cumulative-incidence. Then we simulated 1000 probable epidemic outbreak trajectories and plotted the future daily cumulative incidence predictions based on the estimated time-dependent effective reproduction number. The time-dependent reproduction numbers were estimated using the method proposed by Wallinga & Teunis (Wallinga & Teunis, 2004). We substituted the generation time with the serial interval, which was chosen as a gamma distribution as it accommodates the underlying changing number of events. The mean and standard deviation for the serial interval approximations was 4.4 days and three days, respectively. The shape (number of events in a time-step) and scale (the reciprocal of event rate) of the distribution were 2.15 and 2.04, respectively.

### 3.6. Salient Features of the Dashboard

The components of the dashboard include the daily and cumulative estimates at the national, state, and district levels. These are both the incidence as well as mortality estimates such as deaths per million and case fatality rate, which was defined as the ratio between the number of deaths and the number of diagnosed cases. Other epidemiological parameters such as growth rate, doubling and halving time, as well as other indicators. The dashboard also incorporates other databases such as the sociodemographic information from the Census 2011 population estimates. Test positivity rate, tests per million are also estimated and presented in an intuitive fashion. This dashboard is intended to be used by public health stakeholders. A summary of the intended users and the mode of use has been provided in **Table 2**.

**Interactivity**

The dashboard encourages the user to interact with the data and visualizations and offers a more immersive experience. The dashboard allows the user to select the state of interest and then specify the district that he/she would like to view. The dropdown menu allows for automatic updating based on the state selected. The scale of the axis i.e., linear or logarithmic, can also be specified by the user based on a check box beside the plot. This feature also allows the user to gain better insights into the transmission phenomenon. The three-dimensional graphs also add to the overall interactivity of the dashboard, allowing the user to zoom, pan, and rotate the 3D visualization. The *plotly* library also allows for the districts which are added as traces on the plot to be turned off or on without rendering the whole plot anew, thereby greatly enhancing the user experience.

**Data Visualization**

Data visualization such as epidemic curves, animated plots, time series plots, and 3D graphs have been used to visualize the data. Some of the relationships that were linear on 2D scatterplots show varied non-linear relationship across different geographical regions in the 3D graphs. Also, animated plots using the *ggaminate* package allow for the static plots to come to life while increasing their interpretability.

## RESULTS

The full results of the study could be found as a standalone dashboard in the supplementary appendix. The shape of the epidemic curves not only suggests a classic case of a propagated epidemic but also points to the onset of the second wave of COVID-19 infections in India. Figure 1 shows the epidemic curve (daily and cumulative) for India, which clearly suggest the start of the second wave of COVID-19 infections in India, which is reflected in all the three plots, i.e., a spike in the daily new cases, the increase in cumulative cases and an increase in the Rt has also been noted.

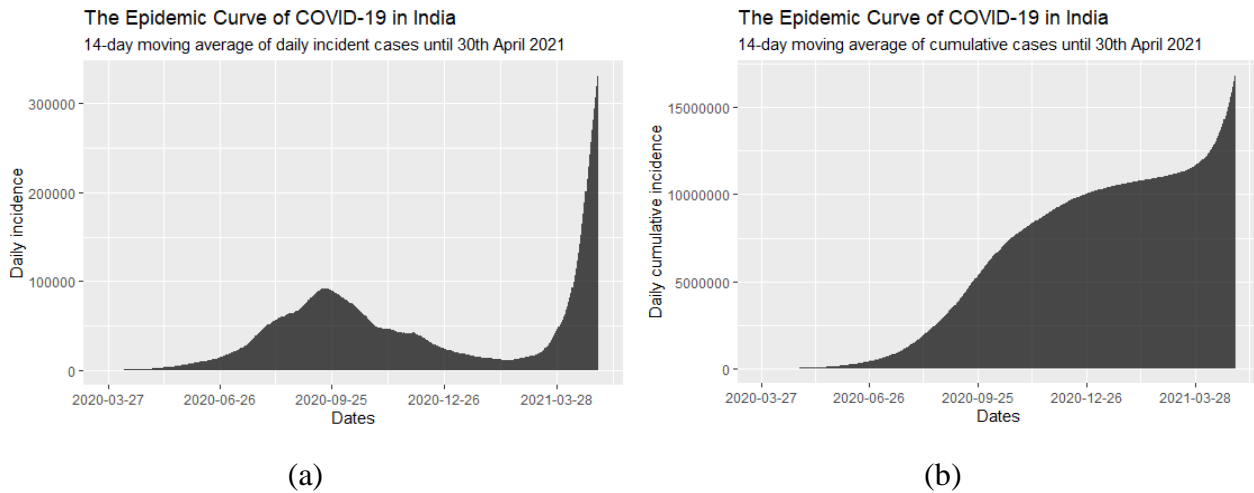

(a)                                                                  (b)

**Figure 1: (a)The epidemic curve of daily new COVID-19 cases in India, (b) Cumulative incidence of COVID-19 in India**

### 1.1. Epidemiological Parameters of the First and Second Wave

Almost all the states in India are experiencing a second wave of COVID-19 infections, as seen in the epidemic curve. The epidemic trajectories suggest a variation in the onset of the second wave among the states. Table 3 summarizes the key epidemiological parameters (growth rate, doubling time, peak of the first wave, and end of the first wave). It was found that the growth rate of the first wave (estimated until the peak of the first wave) is lower than that of the second wave. This was found to be true for almost all the states and union territories of India. Also, the doubling time that was estimated for the second wave was found to be lower than that of the first wave.

### 1.2. Estimation of the time-dependent reproduction number (Rt)

The estimated time-dependent reproduction number (Rt) for India was above the threshold of one (Figure 2.a). Figure 2.b and 2.c illustrate the three-dimensional visualization of the time-dependent reproduction number across different states and districts of Maharashtra, respectively. The findings of the analysis suggest that the R(t) is well above the threshold of one in many states and requires closer monitoring. This also suggests that the second wave of COVID-19 infections in India is still in the exponential phase of the epidemic.

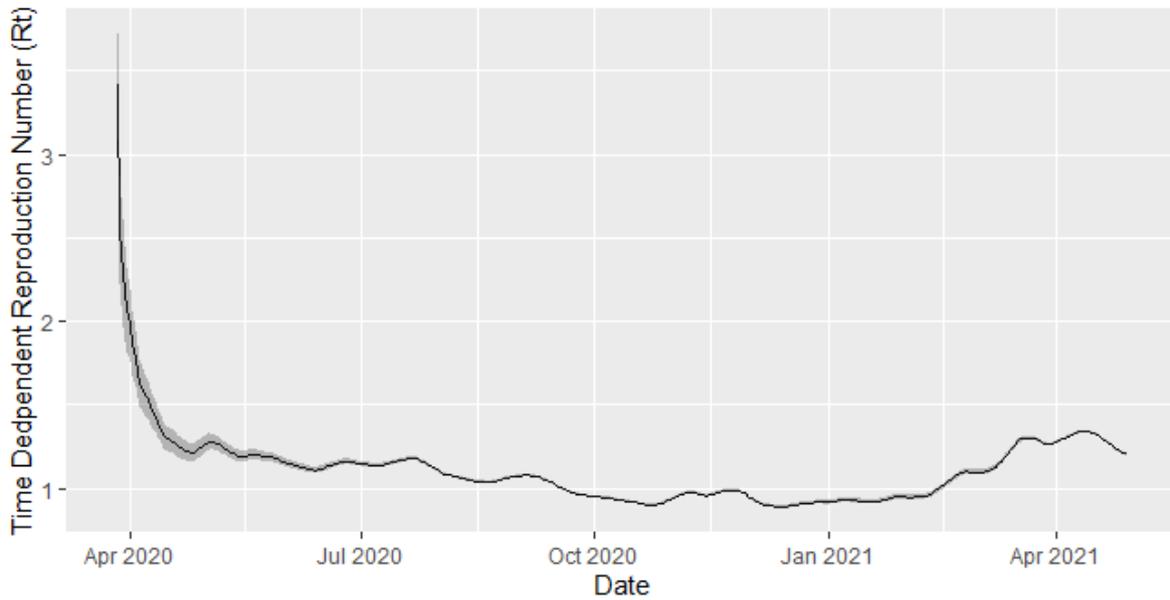

(a)

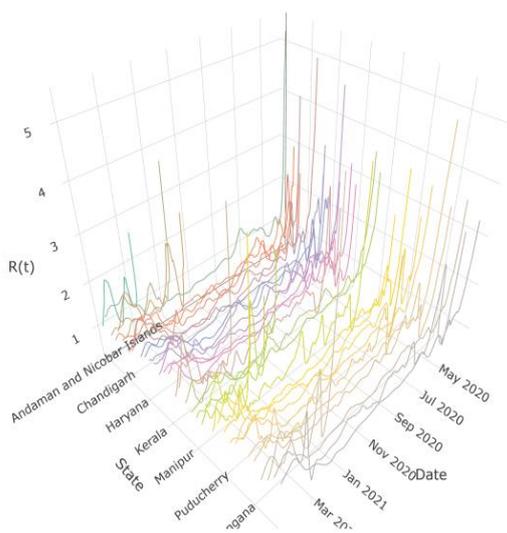 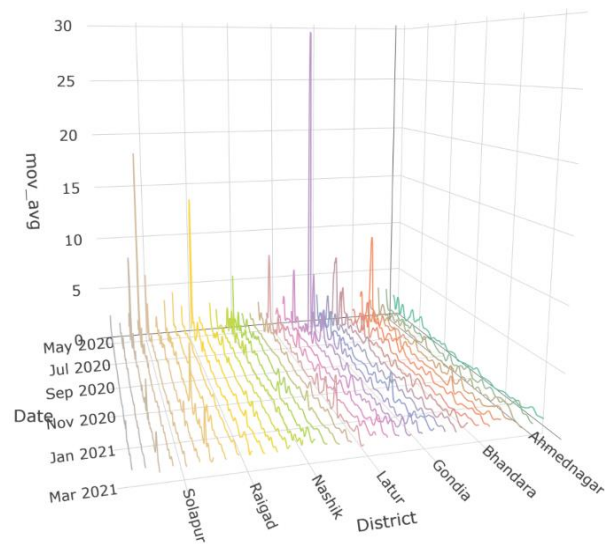

(b)                                                                      (c)

**Figure 2: 2D and 3D visualization of the time-dependent reproduction numbers across different states and districts (Maharashtra for the purpose of illustration)**

### 1.3.     Projection of future daily incidence for the next 15 days

The below figure (Figure 3) plots the projections of the future daily incidence (red-line) based on previous data (red-bars) against the observed daily incidence (blue-bars). It was found that the log-incidence overtime model performs well in predicting the future daily cases of COVID-19 in India. It was also found that since the estimation of projected cases (15$^{th}$ April 2021), there has been about 5% reduction in the R(t), which can be seen in the second panel of the below presented plot. Similar results were found in many of the states of India, especially where the second wave has started early. However, the model did not seem to perform well in states with fewer numbers of cases (daily incidence < 50 cases), especially in the north-eastern states.

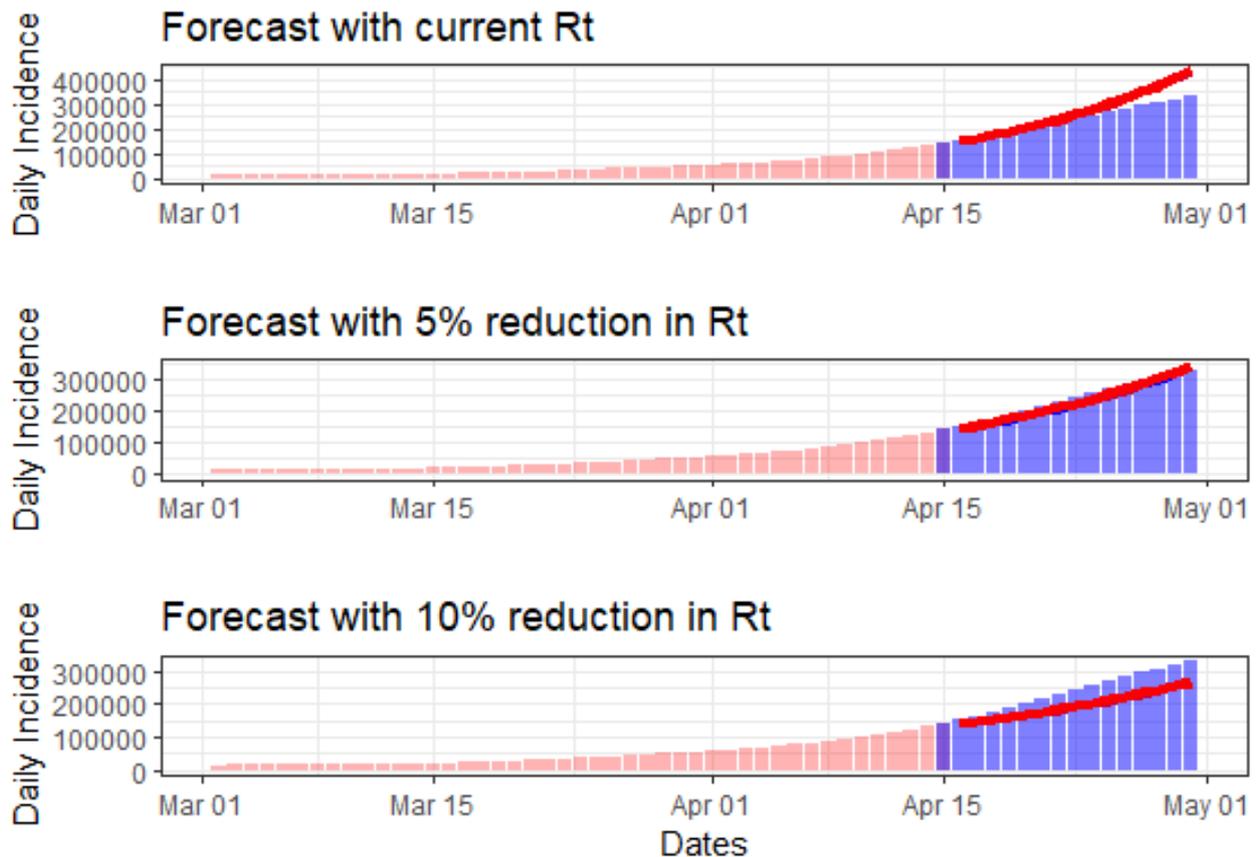

**Figure 3: Projections of future daily incidence from 15th April 2021 to 30th April 2021. The projections are represented as a thick red line, while the observed cases are represented as blue columns on the x-axis. The light red columns on the x-axis represent the previous cases based on which the projections are estimated.**

## DISCUSSION

The current study demonstrates the applications of data science and epidemiologic techniques in tracking and monitoring COVID-19 outbreaks at the district level in India. The dashboard allows the use of scientific methodology and open-source technology to infer from the epidemiologic indicators in an intuitive manner. It not only allows timely information critical for decision making but also provides an interactive interface for the user. Some of the key inferences that can be made from the dashboard tracking COVID-19 epidemic in India are discussed below.

First, the incidence plots suggest the onset of the second wave of COVID-19 infections in India. The growth rate and doubling time have seen a steep increase when looking at the national context. Maharashtra saw a similar trend of increase in the number of cases as well as the key epidemiological parameters in the recent weeks. However, this second wave phenomenon has not yet been observed in all of the states. Also, within Maharashtra, there is a geographical variation across districts in terms of the current magnitude of the epidemic trajectory. This points to the fact that the phenomenon of COVID-19 transmission in India is not homogenous and needs to be looked at in greater detail.

Second, the epidemic trajectory and rate of increase in epidemiological parameters like time-dependent reproduction number, growth rate, and doubling time during the first wave of the COVID-19 epidemic seen last year point to the "flattening of the curve" as suggested by many. In contrast, the rate of increase in the epidemiological parameters at the beginning of the second wave points to the intensity of infections, as seen clearly in the visualization part. Though it may suggest a grim future in terms of the impact of the second wave of COVID-19, this rate of increase may also be reflective of the responsive health system in terms of testing capacity and active case finding. The relationship between case incidence with population density across different districts also is interesting to note. Also, the association between the incidence and mortality estimates reveals that the relationship is not uniformly linear across the country. Different districts, based on their individual attributes, including sociodemographic variables, have an influence on this relationship.

Third, during the first wave, the capacity for testing and availability of tests have been low as compared to the present. The scaling of testing has been possible through the efforts of both governmental agencies and industry professionals. The regulations on price capping have ensured the affordability of COVID-19 testing in India. This has enabled in the improvement of data quality in the present time as compared to the first wave. The scaling of testing is inferred from the visualizations in the dashboard, but we can also see the relationship between the test positivity rate and the number of tests positives. This could mean that as we are increasing testing, the test positivity rate also has increased. Though the raw data required to make this inference is only available for the state of Kerala, this demonstrates the need to look at testing capacity and test positivity as a key performance indicator for the COVID-19 response. As data quality is fundamental to make meaningful decisions relating to public health, it is also important to acknowledge that health information is not immune from the political influences and strategic aspirations of the government.

The projections of the future incidence seem to provide robust estimates of daily incidence over a 15-day time period. The model performs well in estimating the future incidence in most states. The model overestimated the projections in states which saw an early second wave (E.g., Maharashtra, Uttar Pradesh, Chhattisgarh, Delhi). This could also be due to the relative reduction in R(t) owing to the precautionary measures / public health interventions in these states. The model also performs well in all the southern states, such as Andhra Pradesh, Karnataka, Kerala, Tamil Nadu, Telangana, and Puducherry. The model failed to provide robust estimates for the northeastern states as well as some union territories such as Andaman and Nicobar Islands, Ladakh, and Lakshadweep. This may be due to the fact that there was a lack of adequate data and the daily incident cases were less than 50 per day. The model thus has many limitations as it does not take into account many important factors such as health system parameters, testing information, as well as vaccinations.

All of the aspects discussed above are dynamic in nature, i.e., as knowledge on COVID-19 is being updated every day there is an important need to review the situation on a frequent basis. Conventional methods in epidemiology and complex mathematical modeling approaches require high-quality data with little uncertainty, which might be difficult to come by in a low resource setting like India. The collective action of the community in providing support by curating the official information into an information exchange platform for public use is indeed a very welcoming and appreciative initiative. The collective action of various health departments, academic and research institutions, volunteers from the technology industry, and the general public made immense strides in advancing how health information is communicated.

The current dashboard has a lot of benefits over many existing dashboards available for India that have been discussed in the previous sections, and it also has a few limitations. We acknowledge the controversy of whether the data source is credible or not as a complex issue. Though there is a risk of undercounting new cases or deaths, our main data source has been used by many researchers for modeling COVID-19 in India. We also acknowledge that the application in its current form does not take into account the geospatial, genomic, climatic, or factors relating to population dynamics and variation across the district. We hope to incorporate these variables in the subsequent versions of the dashboard. Some of the districts had to be excluded from the analysis due to some empirical constraints. Moreover, the analyses are not free from the biases linked to the source of information provided at https://covid19india.org. Hence, our estimates might not be an accurate representation of the on-ground reality.

# THE ROAD AHEAD

Apart from using the study findings in teaching and academic discussion at various academic and research forums, the authors also intend to use it in the training of district-level program managers. The parent institute of the authors, Sree Chitra Tirunal Institute for Health Sciences and Technology which is under the Department of Science and Technology, has been identified by the Government of India as the Regional Center of Excellence (RoCE) for training in the management of COVID-19 epidemic in India. As part of the training of the district-level program managers of Kerala, the findings of the study would be shared as an interactive dashboard for aiding in decision making. The feedback received would provide insights into user experience and help in the subsequent versions of the dashboard. This could in turn promote data use culture and evidence-informed decision making. We also shared some of the findings of our study with the general public in the form of blog posts to invite citizen participation and public discussion.

# CONCLUSION

This dashboard also demonstrates many possible implications to the district, state, and national level program managers and public health decision-makers. It is a proof of concept that fills a significant gap by presenting a set of tools that are useful for updated analysis and visualization of the COVID-19 epidemiological indicators for India at the district level. It also allows for timely inferences to be made that are region and context-specific in nature rather than having a blanket approach for the whole country.

**Table 2: Summary of some of the COVID-19 Dashboards developed in the Indian context**

| No | Region | Name | Organization | Type | Link | Update | Class | Granularity | Data Visualization | Geospatial Analysis | Data Source |
|----|--------|------|--------------|------|------|--------|-------|-------------|--------------------|---------------------|-------------|
| 1 | India | WHO | Intl' Org. | Interactive |  | Daily | Strategic | National Level | Descriptive | Descriptive | - |
| 2 | India | MoHFW | Govt Org. | Static | https://www.mygov.in/covid-19/ | Daily | Strategic | State Level | No | No | MoHFW |
| 3 | India | ZOHO Analytics | Industry | Interactive | https://www.zoho.com/covid/india/ | Daily | Analytic | State Level | Analytical | Descriptive | - |
| 4 | India | COVID19India.org | Crowd Sourced | Interactive | https://www.covid19india.org/ | Realtime | Analytic | District Level | Analytical | Analytical | Crowd Sorucecd |
| 5 | India | Dataflix | Industry | Interactive | https://www.dataflix.com/covid/india/ | Daily | Analytic | State Level | Descriptive | Descriptive | MoHFW |
| 6 | India | Microsoft Bing | Industry | Interactive | https://www.bing.com/covid/local/india | Daily | Strategic | State Level | Analytical | Descriptive | Multiple |
| 7 | India | National Centre for Disease Control | Govt Org | Interactive | https://ncdc.gov.in/dashboard.php | - | - | - | - | - | - |
| 8 | India | Pracriti - IIT Delhi | Acad. Institution | Interactive | http://pracriti.iitd.ac.in/ | Daily | Analytic | District Level | Analytical | Descriptive | MoHFW |
| 9 | India | United Nations Major Group for Children and Youth | Industry | Static | https://www.asianplanner.com/covid-dashboard | Daily | Strategic | State Level | Analytical | Descriptive | MoHFW |
| 10 | India | Railway Enquiry | Industry | Static | https://railwayenquiry.net/corona | Daily | Strategic | State Level | Descriptive | Descriptive | Multiple |
| 11 | India | National Remote Sensing Centre | Govt Org | Interactive | https://bhuvan-app3.nrsc.gov.in/corona/corona_dashboard/dashboard/dashboard.php?type=citizen | Daily | Strategic | State Level | Descriptive | Descriptive | MoHFW |
| 12 | India | Harvard School of Public Health | Acad. Institution | Interactive | https://www.hsph.harvard.edu/india-center/data-spotlight/ | Daily | Strategic | State Level | Analytical | Descriptive | Multiple |
| 13 | India | FETP Network India - NIE, Chennai | Acad. Institution | Interactive | http://covidindiaupdates.in/ | Daily | Analytic | State Level | Analytical | Descriptive | Multiple |
| 14 | India | National Informatics Centre | Govt. Org | Interactive | https://bharatnetprogress.nic.in/covid19zones/ | Daily | Strategic | District Level | Yes | Descriptive | MoHFW |
| 15 | India | Chandeep | Individual | Interactive | https://www.goodly.co.in/corona-virus-india-state-wise-dashboard/ | Daily | Analytic | State Level | Analytical | Analytical | MoHFW |
| 16 | Nagaland | Govt of Nagaland | Govt Org | Interactive | https://ncdc.gov.in/dashboard.php | Daily | Analytic | National Level | Analytical | No | IDSP, ICMR |
| 17 | Assam | Govt of Assam | Govt Org | Static | https://covid19.assam.gov.in/ | Daily | Strategic | District Level | No | No | Health Dept, Assam, |
| 18 | Karnataka | Govt of Karnataka | Govt Org | Interactive | https://covid19.karnataka.gov.in/covid-dashboard/dashboard.html | Daily | Analytic | District Level | Yes | Descriptive | Health Dept, Karnataka |
| 19 | Odisha | Dept of Health and Family Welfare | Govt Org | Static | https://health.odisha.gov.in/covid19-dashboard.html | Daily | Strategic | District Level | Descriptive | Descriptive | Health Dept, Odisha |
| 20 | Punjab | Drona Maps | Industry | Interactive | https://dronamaps.com/m.corona.html | Daily | Analytic | District Level | Analytical | Analytical | - |
| 21 | Bengaluru | Greater Bengaluru Municipal Corporation | Govt Org | Interactive | https://analysis.bbmpgov.in/#/dashboards/7WGKgnEBHZrt-aeeDTHg?embedd=1&toolbar=0&t=bbmp.com | Daily | Analytic | Ward Level | Analytical | Analytical | Health Dept BBMP |
| 22 | Delhi | Govt of Delhi | Govt Org | Static | https://coronabeds.jantasamvad.org/ | Daily | Strategic | State Level | No | No | - |
| 23 | Punjab | Govt of Punjab | Govt Org | Interactive | https://corona.punjab.gov.in/ | Daily | Strategic | District Level | Descriptive | Descriptive | MoHFW |
| 24 | Puducherry | Department of Revenue & Disaster | Govt Org | Interactive | https://covid19dashboard.py.gov.in/ | Daily | Strategic | State Level | Analytical | No | - |

| | | Management, Puducherry | | | | | | | | | |
|---|---|---|---|---|---|---|---|---|---|---|---|
| 25 | Kerala | Center for Development of Imaging Technology, Govt of Kerala | Govt Org | Interactive | https://dashboard.kerala.gov.in/index.php | Daily | Analytic | District Level | Analytical | Analytical | - |
| 26 | Himachal Pradesh | Govt of Himachal Pradesh | Govt Org | Static | http://covidportal.hp.gov.in/ | Daily | Strategic | State Level | No | No | - |
| 27 | Odisha | Govt of Odisha | Govt Org | Interactive | https://statedashboard.odisha.gov.in/ | Daily | Analytic | District Level | Analytical | Analytical | - |
| 28 | Uttar Pradesh | Govt of Odisha | Govt Org | Interactive | https://updgmh-covid19.maps.arcgis.com/apps/opsdashboard/index.html#/bfc888151feb48928a4f6885ca20e83c | Daily | Strategic | District Level | Descriptive | Descriptive | - |
| 29 | India | IIT Madras | Acad. Institution | Interactive | http://www.ae.iitm.ac.in/~parama/covid19/India/covid19India.html | Daily | Analytic | State Level | Analytical | No | covid19india.org |
| 30 | India | National Informatics Centre | Govt Org | Static | https://www.mygov.in/covid-19/ | Daily | Strategic | State Level | No | No | MoHFW |
| 31 | India | Covid Tracking India | Acad. Institution | Interactive | https://covidtracking.in/ | Daily | Analytic | State Level | Yes | Analytical | covid19india.org |

**Table 2: Summary of the intended users of the data analysis**

| | Intended Users | Intended Purpose | Mode | Updates |
|---|---|---|---|---|
| 1 | District-level program managers | Evidence informed decision making; feedback on future improvements | Interactive Dashboard | Real-time |
| 2 | Decision makers | Evidence informed decision making | Dashboard | Periodic |
| 3 | Academia and researchers | discussion on methods; peer-review; academic discourse | Dashboard + Methodology | Real-time |
| 4 | General public | Citizen involvement, public discourse, media and information professionals | Website / Blog | Periodic |

**Table 3: The epidemiological indicators of the first wave and second wave across different states of India**

| No | State | First Wave Peak | Start of Second Wave | Growth Rate | | Doubling Time | |
|---|---|---|---|---|---|---|---|
| | | | | First Wave | Second Wave | First Wave | Second Wave |
| 1 | Andaman and Nicobar Islands | 15-08-2020 | 30-01-2021 | 0.038 | 0.057 | 18.2 | 12.2 |
| 2 | Andhra Pradesh | 08-09-2020 | 20-02-2021 | 0.049 | 0.083 | 14.1 | 8.3 |
| 3 | Arunachal Pradesh | 02-10-2020 | 04-02-2021 | 0.039 | 0.172 | 17.6 | 4 |
| 4 | Assam | 10-09-2020 | 14-02-2021 | 0.056 | 0.075 | 12.5 | 9.3 |
| 5 | Bihar | 19-08-2020 | 12-03-2021 | 0.054 | 0.128 | 12.8 | 5.4 |
| 6 | Chandigarh | 20-09-2020 | 15-02-2021 | 0.032 | 0.047 | 22 | 14.7 |
| 7 | Chhattisgarh | 28-09-2020 | 04-03-2021 | 0.051 | 0.083 | 13.5 | 8.4 |
| 8 | Delhi | 14-11-2020 | 18-02-2021 | 0.019 | 0.084 | 35.8 | 8.2 |
| 9 | Goa | 22-09-2020 | 24-02-2021 | 0.044 | 0.06 | 15.9 | 11.5 |
| 10 | Gujarat | 04-12-2020 | 18-02-2021 | 0.014 | 0.059 | 50 | 11.7 |
| 11 | Haryana | 23-11-2020 | 14-02-2021 | 0.027 | 0.069 | 25.6 | 10.1 |
| 12 | Himachal Pradesh | 07-12-2020 | 24-02-2021 | 0.026 | 0.066 | 26.9 | 10.5 |
| 13 | Jammu and Kashmir | 21-09-2020 | 10-02-2021 | 0.031 | 0.052 | 22.3 | 13.4 |
| 14 | Jharkhand | 09-09-2020 | 24-02-2021 | 0.048 | 0.088 | 14.5 | 7.9 |
| 15 | Karnataka | 12-10-2020 | 24-02-2021 | 0.044 | 0.072 | 15.8 | 9.7 |
| 16 | Kerala | 13-10-2020 | 25-03-2021 | 0.041 | 0.08 | 17 | 8.6 |
| 17 | Ladakh | 16-11-2020 | 11-03-2021 | 0.019 | 0.114 | 36.2 | 6.1 |
| 18 | Lakshadweep | 13-02-2021 | 05-04-2021 | 0.061 | 0.153 | 11.4 | 4.5 |
| 19 | Madhya Pradesh | 24-09-2020 | 13-02-2021 | 0.026 | 0.062 | 26.7 | 11.2 |
| 20 | Maharashtra | 18-09-2020 | 11-02-2021 | 0.033 | 0.047 | 20.7 | 14.8 |
| 21 | Manipur | 23-10-2020 | 13-03-2021 | 0.026 | 0.095 | 27 | 7.3 |
| 22 | Meghalaya | 12-10-2020 | 27-02-2021 | 0.042 | 0.093 | 16.7 | 7.5 |
| 23 | Mizoram | 14-11-2020 | 13-03-2021 | 0.016 | 0.112 | 43 | 6.2 |
| 24 | Nagaland | 10-08-2020 | 28-02-2021 | 0.046 | 0.087 | 14.9 | 8 |
| 25 | Odisha | 27-09-2020 | 24-02-2021 | 0.047 | 0.079 | 14.6 | 8.8 |
| 26 | Puducherry | 02-10-2020 | 23-02-2021 | 0.045 | 0.067 | 15.3 | 10.3 |
| 27 | Punjab | 21-09-2020 | 28-01-2021 | 0.037 | 0.042 | 18.8 | 16.6 |
| 28 | Rajasthan | 01-12-2020 | 24-02-2021 | 0.019 | 0.084 | 36.6 | 8.3 |
| 29 | Sikkim | 27-09-2020 | 15-02-2021 | 0.031 | 0.066 | 22.4 | 10.6 |
| 30 | Tamil Nadu | 04-08-2020 | 25-02-2021 | 0.05 | 0.061 | 13.9 | 11.4 |
| 31 | Telangana | 06-09-2020 | 23-02-2021 | 0.04 | 0.065 | 17.5 | 10.6 |
| 32 | Tripura | 13-09-2020 | 04-02-2021 | 0.035 | 0.063 | 19.7 | 10.9 |
| 33 | Uttar Pradesh | 18-09-2020 | 22-02-2021 | 0.038 | 0.105 | 18.5 | 6.6 |

| 34 | Uttarakhand | 22-09-2020 | 01-03-2021 | 0.045 | 0.093 | 15.6 | 7.5 |
| 35 | West Bengal | 31-10-2020 | 21-02-2021 | 0.032 | 0.072 | 21.5 | 9.6 |

## REFERENCES AND CITATIONS